\documentclass[fleqn,usenatbib]{mnras}

\usepackage{newtxtext,newtxmath}
\usepackage[T1]{fontenc}

\DeclareRobustCommand{\VAN}[3]{#2}
\let\VANthebibliography\thebibliography
\def\thebibliography{\DeclareRobustCommand{\VAN}[3]{##3}\VANthebibliography}

\usepackage{graphicx}	% Including figure files
\usepackage{amsmath}	% Advanced maths commands
\usepackage{placeins}
\usepackage{natbib}
\usepackage{hyperref}
\usepackage{placeins}
\hypersetup{
    colorlinks=true,
    linkcolor=blue,}
\newcommand\sbullet[1][.5]{\mathbin{\vcenter{\hbox{\scalebox{#1}{$\bullet$}}}}}
\newcommand{\ha}{$\rm H\,\alpha$}
\newcommand{\igd}{$i_{\rm GD}$}
\newcommand{\iha}{$i_{\rm H\alpha}$}
\newcommand{\inpoi}{$i_{\rm NPOI}$}
\title[Inclination Angles for Be Stars Determined Using Machine Learning]{Inclination Angles for Be Stars Determined Using Machine Learning}
\author[Lailey \& Sigut]{
B. D. Lailey$^{1}$\thanks{E-mail: blailey2@uwo.ca} and
T.\ A. A. Sigut$^{1,2}$
\\
$^{1}$Department of Physics and Astronomy, The University of Western Ontario, 1151 Richmond Street, London N6A 3K7, Canada \\
$^{2}$Institute for Earth and Space Exploration (IESX), The University of Western Ontario, Canada
}

\date{Accepted 2023 October 24. Received 2023 October 24; in original form 2023 June 30}

\pubyear{2023}

\begin{document}
\label{firstpage}
\pagerange{\pageref{firstpage}--\pageref{lastpage}}
\maketitle

\begin{abstract}
We test the viability of training machine learning algorithms with synthetic \ha\ line profiles to determine the inclination angles of Be stars (the angle between the central B~star's rotation axis and the observer's line of sight) from a single observed medium-resolution, moderate S/N, spectrum. The performance of three different machine learning algorithms were compared: neural networks tasked with regression, neural networks tasked with classification, and support vector regression. Of these three algorithms, neural networks tasked with regression consistently outperformed the other methods with a RMSE error of $7.6^\circ$ on an observational sample of 92 galactic Be stars with inclination angles known from direct \ha\ profile fitting, from the spectroscopic signature of gravitational darkening, and, in a few cases, from interferometric observations that resolved the disk. The trained neural networks enable a quick and useful determination of the inclination angles of observed Be stars which can be used to search for correlated spin axes in young open clusters or to extract an equatorial rotation velocity from a measurement of $v\sin i$. 

%We test the viability of training machine learning algorithms to determine the inclination angles of Be stars from a single medium resolution observed spectrum in the vicinity of $\rm H\,\alpha$. The algorithms, which were trained on synthetic spectra, could reliably determine the inclination angles of observed Be stars. We trained and compared three different machine learning algorithms: neural networks tasked with regression, neural networks tasked with classification, and support vector regression. Of the three machine learning algorithms that we trained, neural networks tasked with regression consistently outperform the other two; on an observational sample of 92 galactic Be stars, with inclination angles determined using $\rm H\,\alpha$ profile fitting, they have a root mean squared error of $7.6^\circ$ and in a case study of 11 nearby Be stars, with inclination angles determined from long baseline optical interferometry, they have a root mean squared error of $12.3^\circ$. The trained algorithms enable a quick, useful determination of the inclination angles of observed Be stars which can be used to search for correlated spin axes in young open clusters or to extract an equatorial rotation velocity from a measurement of $v\sin i$. 

\end{abstract}


%%%%%%%%%%%%%%%%%%%%%%%%%%%%%%%%%%%%%%%%%%%%%%%%%%

%%%%%%%%%%%%%%%%% BODY OF PAPER %%%%%%%%%%%%%%%%%%

\section{Introduction}
\label{section:1}

\subsection{Machine Learning in Astronomy}
Astronomers are increasingly turning to machine learning to provide automated detection, analysis, and classification in response to large scale surveys that produce unprecedentedly large datasets \citep{Baron2019}. Machine learning differs from traditional model-fitting techniques in that the model is constructed according to the input data rather than being predefined \citep{Ivezic2020}. The flexible nature of machine learning algorithms make them suited to a wide variety of tasks. In astronomical research, common uses of machine learning include classifying objects of interest from large databases \citep{Sanchez2018, Wang2022}, dimensionality reduction \citep{Portillo2020, Kovacevic2022}, anomaly detection \citep{Baron2017, Giles2020}, building models that use more parameters than is possible with classical models \citep{Huertas-Company2008}, and visualizing datasets with a high number of parameters \citep{Giles2018, Reis2021}.

Broadly speaking, machine learning can be divided into supervised and unsupervised algorithms. In supervised machine learning, a set of input features are mapped to a target variable based on labels provided by a human expert \citep{Ivezic2020}. In unsupervised machine learning, labels are not included and the algorithms are frequently used to cluster data into groups, reduce dimensionality, and detect anomalies \citep{Baron2019}. 

\subsection{Machine Learning in Be Star Research}
Classical Be stars are rapidly-rotating, B-type, main sequence stars that are surrounded by an equatorial, circumstellar, decretion disc \citep{Porter2003}. The defining characteristic of a Be star is the presence of emission in the hydrogen Balmer series, notably \ha, owing to the presence of the disc \citep{Slettebak1982}. The exact mechanism that puts the disc gas into orbit is unknown, but it is thought to be related to near critical rotation, perhaps driven by the redistribution of angular momentum within the star \citep{Granada2013, Rivinius2013}.   

Machine learning has emerged as a promising technique to identify Be star candidates in databases produced by large photometric and spectroscopic surveys. \cite{Bromova2014} used wavelet transformations to reduce the dimensionality of approximately 2,300 spectra of about 300 Be and B[e] stars in the vicinity of \ha. Each spectrum was given a label corresponding to pure emission, absorption smaller than $1/3$ of the emission peak, absorption greater than $1/3$ of the emission peak, and no emission. These labels were then used to train a support vector machine \citep{Vapnik1999} to classify the spectra into emission stars and normal stars. Although \cite{Bromova2014} were not explicitly concerned with the determination of inclination angles, their approach shares significant similarities with the present work. 

\cite{Reis2018} searched dr14 of the APOGEE near-infrared survey using methods based on anomaly detection. A random forest algorithm \citep{Ho1995} was trained, using a sample consisting of both synthetic and observed spectra, to create a matrix of similarity scores between each pair of spectra based on the likelihood that a given pair would end up in the same terminal branch of the random forest. The similarity matrix was then used as the input for a t-SNE algorithm \citep{vanderMaaten2008} to reduce dimensionality and help with visualization. The spectra with the lowest similarity scores and their nearest neighbors were then manually inspected yielding (among other finds) 40 previously undiscovered, classical Be stars.          

\cite{Wang2022} found 1,162 Be star candidates in dr7 of the LAMOST survey by searching for \ha\ emission in the spectra of early type stars using the ResNet convolutional neural network \citep{He2015}, combined with a series of tests to remove confounding objects such as B[e] and Herbig stars. A follow up series of tests on the Be star candidates yielded 183 previously undiscovered classical Be stars. 

%The present work uses supervised machine learning algorithms trained on synthetic spectra to determine Be star inclination angles from a single medium-to-high resolution spectrum in the vicinity of $\rm H\,\alpha$. This is accomplished by automating the method of \citet[see Section \ref{inc_angle}]{Sigut2020} using three different implementations of supervised machine learning: support vector regression, neural networks tasked with regression, and neural networks tasked with classification. The results of each of the three implementations will then be compared across two samples of observed Be stars.

The present work seeks to extend machine learning as applied to the Be stars to include the automatic determination of quantitative information from their spectra. As it is well known that the morphology of the \ha\ line strongly reflects how the star-disk system is viewed (see Figure~\ref{fig:iexample} and the discussion below), we target the extraction of the stellar viewing inclination of the central star from a single, continuum normalized spectrum of a moderate resolution centred on \ha. The performance of three supervised machine learning algorithms, each trained on synthetic spectra, are compared: neural networks tasked with regression, neural networks tasked with classification, and support vector regression. Each algorithm is then applied to an observed sample of Be star \ha\ spectra to judge performance in realistic cases.

\subsection{The inclination angle and its relationship to \texorpdfstring{H$\alpha$}{Ha} morphology}
\label{inc_angle}
\begin{figure}
\centering
\includegraphics[width=0.47\textwidth]{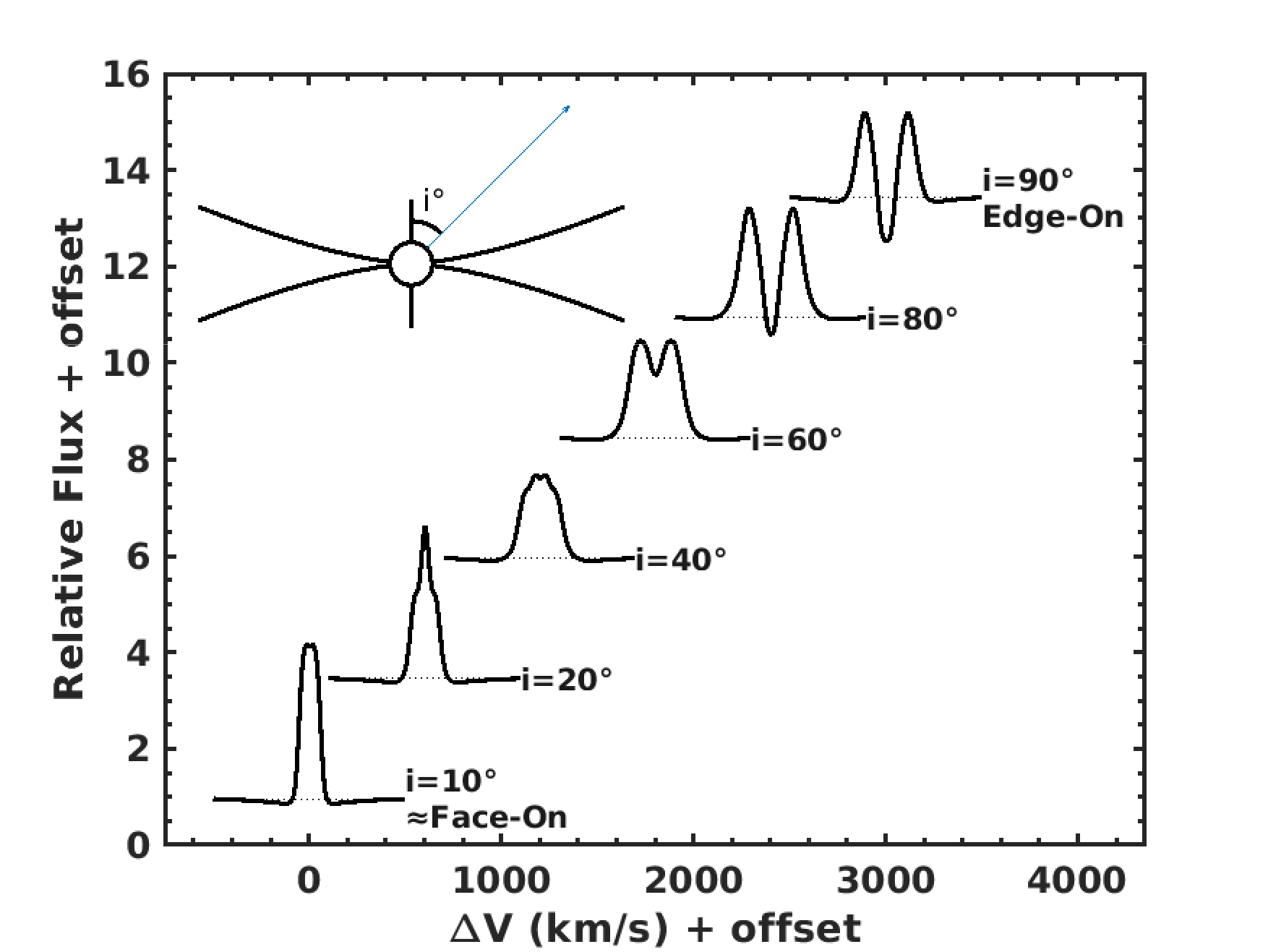}
\caption{\ha\ emission line profile computed by \texttt{Beray} for a $\rm 5\,M_{\odot}$ Be star at a resolution of $\mathcal{R} = 10,000$ viewed at the indicated inclination angles. The inclination angle is the angle between the star's rotation axis and the observer's line of sight as illustrated by the upper-left insert. Here the blue arrow points to the distant observer. Note the strong change in the line profile shape as the viewing angle goes from $i=10^\circ$ (a nearly face-on disc) to $i=90^\circ$ (an edge-on disc). The disc density parameters $\log\rho_0=-10.1$, $n=3.0$, and $R_{\rm d}=65\,R_*$ (See Eq.~\ref{eq:rho}) were used in the \texttt{Beray} calculation.}
\label{fig:iexample}
\end{figure}

The inclination angle, \textit{i}, is the angle between a star's axis of rotation and an observer's line of sight and ranges from $0^{\circ}$ to $90^{\circ}$ for pole-on and edge-on observations respectively (see Figure~\ref{fig:iexample}). It is usually assumed that stellar rotation axes are randomly oriented in space which leads to an expected $p(i)\,di=\sin i\,di$ distribution for any observed sample of stars \citep{gray2021}. 

\cite{Corsaro_2017} cast doubt on the assumption of random inclinations by finding significant spin axis alignment for the red giant stars in the old open clusters NGC 6791 and NGC 6819 using asteroseismology. \cite{Corsaro_2017} investigated 48 oscillating red giant stars with masses in the range of 1.1--1.7 $\rm M_\odot$ and found that about $70$ percent of the stars in each cluster showed a strong level of alignment. The probability that these alignments arose by chance from an underlying random distribution was calculated to be below $10^{-7}$ for NGC 6819 and below $10^{-9}$ for NGC 6791 \citep{Corsaro_2017}. Conversely, the inclination angle distribution obtained from a sample of 36 field red giants showed no significant spin alignment \citep{Corsaro_2017}. Hydrodynamical simulations \citep{Corsaro_2017} and numerical simulations of the effects of shear versus compressive turbulence \citep{Rey-Raposo2018} suggest that if a significant fraction of a star cluster's initial kinetic energy is rotational, then stars can form in a cluster with significant correlations in the direction of their rotation axes that can persist over Gyr timescales.  

The strongly correlated spin alignments found by \cite{Corsaro_2017} have been contested by \cite{Mosser2018} and \cite{Gehan2020}, who attributed them to a combination of systematic bias that favoured low inclination angles and neglecting to account for the impossibility of measuring inclination angles near either $0^\circ$ or $90^\circ$. A re-analysis by \cite{Mosser2018} of the spin alignments of both NGC 6819 and NGC 6791 found the inclination angle distribution of both open clusters to be consistent with a $\sin i$ distribution upon taking these effects into account. \cite{Gehan2020}'s analysis supported \cite{Corsaro_2017}'s conclusion that the distribution of the field red giant stars was isotropic, but was unable to test the conclusions on spin-alignment in open clusters because their method is unsuitable for red clump stars. \cite{Gehan2020} urged caution in accepting strongly aligned stellar spins in open clusters and highlighted the need for a dedicated study using another method.     

Be stars offer an alternative avenue to search for correlated spin axes in young open clusters. This is because Be stars are bright, common ($\approx 20$ percent of main sequence B stars are Be stars \citep{Zorec1997}), and their inclination angles can be reliably determined spectroscopically (see below). Also, for bright and nearby Be stars, $i$ can be reliably determined using long baseline optical interferometry (LBOI) observations of the star-disc system \citep{ vanBelle2012}. Additionally, there are methods based on gravitational darkening, in which rapid rotation causes the stellar intensity to vary with latitude \citep{vonZeipel1924, Collins1963}, and $i$ is extracted from detailed spectral synthesis \citep{Townsend2004,Fremat2005,Zorec2016}. 

\cite{Sigut2020} showed that spectral synthesis of \ha\ can accurately determine the orientation of a Be star's disc, and as the disc is in the star's equatorial plane, the inclination of the star itself. The method of \cite{Sigut2020} leverages the fact that the morphology of a Be star's \ha\ emission-line profile varies strongly with inclination even if the disc size and density structure is held constant. This is shown in Figure \ref{fig:iexample}; here low inclinations give rise to singly-peaked emission in \ha, moderate inclinations result in doubly-peaked emission, and high inclinations result in doubly-peaked lines with deep shell absorption \citep{Porter2003}. By comparing a single observed \ha\ profile to a library of synthetic spectra computed using the \texttt{Bedisk} and \texttt{Beray} suite of codes \citep{Sigut2018}, \cite{Sigut2020} were able to recover the inclination angles of 11 Be stars to within $\pm 10^{\circ}$ as compared to LBOI determined inclinations. \cite{Sigut2022} further test the H$\alpha$ technique using a sample of Be stars with inclinations available from gravitational-darkening studies \citep{Zorec2016} and find good agreement between the two methods.

\subsection{Organization}

Section~\ref{section:2} describes the synthetic Be star spectra used to train the machine learning algorithms. Section~\ref{Algorithms} describes the three machine learning algorithms and the associated performance metrics by which they have been evaluated. Section~\ref{section:4} details the procedure for optimizing user-defined model parameters, called hyper-parameters, that must be tuned for each algorithm in order to ensure optimal performance and discusses the accuracy achieved in the synthetic test samples. The results of testing the trained algorithms on observed \ha\ profiles for a sample of 92 Be stars, with available inclination angle determinations from gravity darkening \citep{Zorec2016} and \ha\ profile fitting \citep{Sigut2022}, are found in Section~\ref{section:zorec}. Section~\ref{section:6} contains a case study of using the trained algorithms on 11 nearby Be stars with well-constrained inclination angle determinations from LBOI. A discussion of our results follows in Section~\ref{section:conclusion}.   

\section{Synthetic Training Spectra}
\label{section:2}
In order to train machine learning algorithms to determine the inclination angles of Be stars, large libraries of synthetic spectra were generated centred on the vacuum value of \ha, $\rm 6564.6 \,$\AA. Each individual model \ha\ profile is represented by 201 continuum-normalized flux values covering the region $\rm \pm 1000$ $\rm km\,s^{-1}$ from line centre. One library of \ha\ line profiles, corresponding to a range of equatorial disc density models, was generated for each of the central B star masses given in Table \ref{stellar_properties}, which correspond to spectral types ranging from approximately B9V to B0.5V. 

\subsection{Creating the libraries of synthetic spectra}
\label{spectra_libraries}
The libraries of Be star \ha\ line profiles were computed by \citet{Sigut2020} using the \texttt{Bedisk} and \texttt{Beray} suite of codes \citep{Sigut2018}. \citet{Ekstrom2012}'s stellar evolutionary models for a core hydrogen fraction of $X = 0.3$, which corresponds approximately to the middle-age main sequence, were used to generate the radii, luminosities, and effective temperatures of the central B stars. Table~\ref{stellar_properties} details the stellar properties adopted.

\begin{table}
\centering
\begin{tabular}{c c c c c} 
\hline
Mass  & Radius  & Luminosity  & $T_{eff}$  & Log(g) \\ [0.49ex] 
($\rm M_{\odot}$) & $\rm (R_{\odot})$ & ($\rm L_{\odot}$) & (K) & \\
\hline
3.00 & 2.9 & 1.12e+02 & 11,000 & 4.0 \\ 
4.00 & 3.5 & 3.44e+02 & 13,400 & 4.0  \\ 
5.00 & 3.9 & 7.93e+02 & 15,600 & 4.0\\ 
6.00 & 4.3 & 1.52e+03 & 17,400 & 3.9 \\ 
7.00 & 4.7 & 2.65e+03 & 19,200 & 3.9 \\ 
8.00 & 5.1 & 4.23e+03 & 20,600 & 3.9 \\ 
9.00 & 5.4 & 6.28e+03 & 22,000 & 3.9 \\ 
10.0 & 5.7 & 8.88e+03 & 23,400 & 3.9 \\ 
12.0 & 6.4 & 1.58e+04 & 25,600 & 3.9 \\ 
13.9 & 7.0 & 2.51e+04 & 27,400 & 3.9 \\ 
15.9 & 7.7 & 3.69e+04 & 29,000 & 3.9  \\ \hline
\end{tabular}
\caption{Stellar properties adopted from \citet{Ekstrom2012} corresponding to a core hydrogen fraction of $X=0.3$.}
\label{stellar_properties}
\end{table}

\texttt{Bedisk} outputs the radiative equilibrium temperatures in the Be star's circumstellar disc given the central B star's photoionizing radiation field and density structure as inputs \citep{Sigut2007}. If the distance from the rotation axis of the central B star is $R$, the central B star's radius is $R_*$, the distance above the equatorial plane is $Z$, and the disc scale height is $H$, the density structure of the disc is parameterized by
\begin{equation}
\label{eq:rho}
%\centering
\rho(R,Z) = \rho_0 \left(\frac{R_*}{R}\right)^n e^{-(\frac{Z}{H})^2},
\end{equation}
where $\rho_0$ and $n$ are free parameters that can be adjusted to match observations.

The scale height, H, of a disc in vertical hydrostatic equilibrium is given by 
\begin{equation}
%\centering
H = \left[\frac{c_{\rm s}(T_0)}{V_{\rm K}(R)}\right]\,R \;,  
\end{equation}
where temperature $T_0 = 0.6\,T_{\rm eff}$, $c_{\rm s}$ is the speed of sound at $T_0$, and $V_{\rm K}(R)$ is the Keplerian orbital speed at distance $R$ \citep[see][]{Sigut2020}.

For each central B star mass given in Table \ref{stellar_properties}, 165 different discs were considered, comprised of 15 values of $\rho_{\rm 0}$ distributed evenly in log-space between $\rm 10^{-12} \,g\, cm^{-3}$ and $\rm 10^{-10} \,g\, cm^{-3}$ and 11 values of $n$ between 1.5 and 4 in increments of 0.25 \citep{Sigut2020}. A \texttt{Bedisk} model was computed for each of the 165 permutations and then the hydrogen level populations computed by \texttt{Bedisk} were used by \texttt{Beray} to compute individual $\rm H\,\alpha$ line profiles. \texttt{Beray} accomplishes this task by solving the radiative transfer equation along a series of rays directed at the observer \citep{sigut2010,Sigut2018}. The composite disc-plus-star \ha\ profile is computed in a unified way by incorporating the relevant boundary condition for each ray. Rays that terminate on the stellar surface use a Doppler-shifted, photospheric \ha\ profile for the upwind boundary; rays that pass through the disc but miss the star assume no incident radiation. This allows the computed profiles to be directly compared with observed profiles (after convolution to the correct spectral resolution). Calculating the \ha\ line profiles introduces two new parameters, $R_{\rm D}$, which is the outer radius of the disc and $i$, which is the inclination. Seven disc sizes, from 5 $R_*$ to 65 $R_*$ in steps of 10 $R_*$, and ten inclinations, from $0^\circ$ to $90^\circ$ in steps of $10^\circ$, were considered. Each of the 11 central Be star masses detailed in Table \ref{stellar_properties} has an associated library containing 11,550 line profiles resulting in 127,050 \ha\ line profiles overall. 

\subsection{Samples of synthetic spectra}
\label{samples_of_spectra}

Several different samples of \ha\ spectra are used in this work, and the following naming conventions are employed. Previously in Section~\ref{spectra_libraries}, a {\em library} of 11,550 synthetic spectra was created for each stellar mass in Table \ref{stellar_properties}. This current section details the creation of a {\em sample} of $\sim$8,000 synthetic spectra from each of the libraries of synthetic spectra. Section \ref{train,val,test} describes how these samples of synthetic spectra are further divided into {\em training, validation, and test sets}. The training, validation, and test sets are used to optimize the algorithms' hyper-parameters in Section \ref{section:4} and to train the algorithms in Section \ref{section:training}. Once trained, the algorithms will be used to determine the inclination angles of two samples of observed spectra: the 92 star {\em Zorec sample} in Section \ref{section:zorec} and the 11 star {\em NPOI sample} in Section \ref{section:6}.  

To create a sample of synthetic spectra from a profile library, the desired number of spectra, $n_{\rm spec}$, is specified. Then, only \ha\ spectra that have an average, absolute percentage difference from the reference photospheric profile (for the same mass) of 3 percent or more are selected randomly from the line profile library corresponding to a central B star of a given mass. Profiles too similar to the reference photospheric line profile are not included because they lack significant line emission (or shell absorption), and therefore poorly constrain the inclination angle. As line profiles within this 3 percent threshold are excluded from the sample, it is not possible to use all 11,550 \ha\ line profiles contained within a given library. This work uses a sample of $\sim$8,000 \ha\ line profiles for each central B star mass.    

Two additional parameters that need to be specified when creating a sample are the spectral resolution, $\mathcal{R}$, and the signal to noise ratio, $\rm S/N$. If $\Delta\lambda$ is the characteristic width of the instrumental profile, then the resolution of the spectra is defined as $\mathcal{R}\equiv\lambda/\Delta\lambda$. The signal to noise ratio is the ratio between the measured flux of the signal to that of the noise in the continuum adjacent to the line, i.e. $\rm S/N = 100$ spectra will have 1$\,\sigma$ error bar magnitudes equal to 1 percent of their corresponding flux measurements. The profiles were generated at $\mathcal{R} = 10,000$ and $\rm S/N = 25$. The resolution was chosen because it matches that of the Zorec and NPOI samples of observed spectra in Section \ref{section:6}. 

Although the observed sample spectra have $\rm S/N \gtrsim 100$, initial testing found that algorithms trained on $\rm S/N = 25$ profiles outperformed algorithms trained on $\rm S/N = 100$ profiles at predicting the inclination angles of observed Be stars, possibly because the algorithms trained at $\rm S/N = 100$ were overspecialized to synthetic profiles and could not deal effectively with the deviations from those profiles exhibited by observed spectra.

\begin{figure}
\centering
\includegraphics[width=0.47\textwidth]{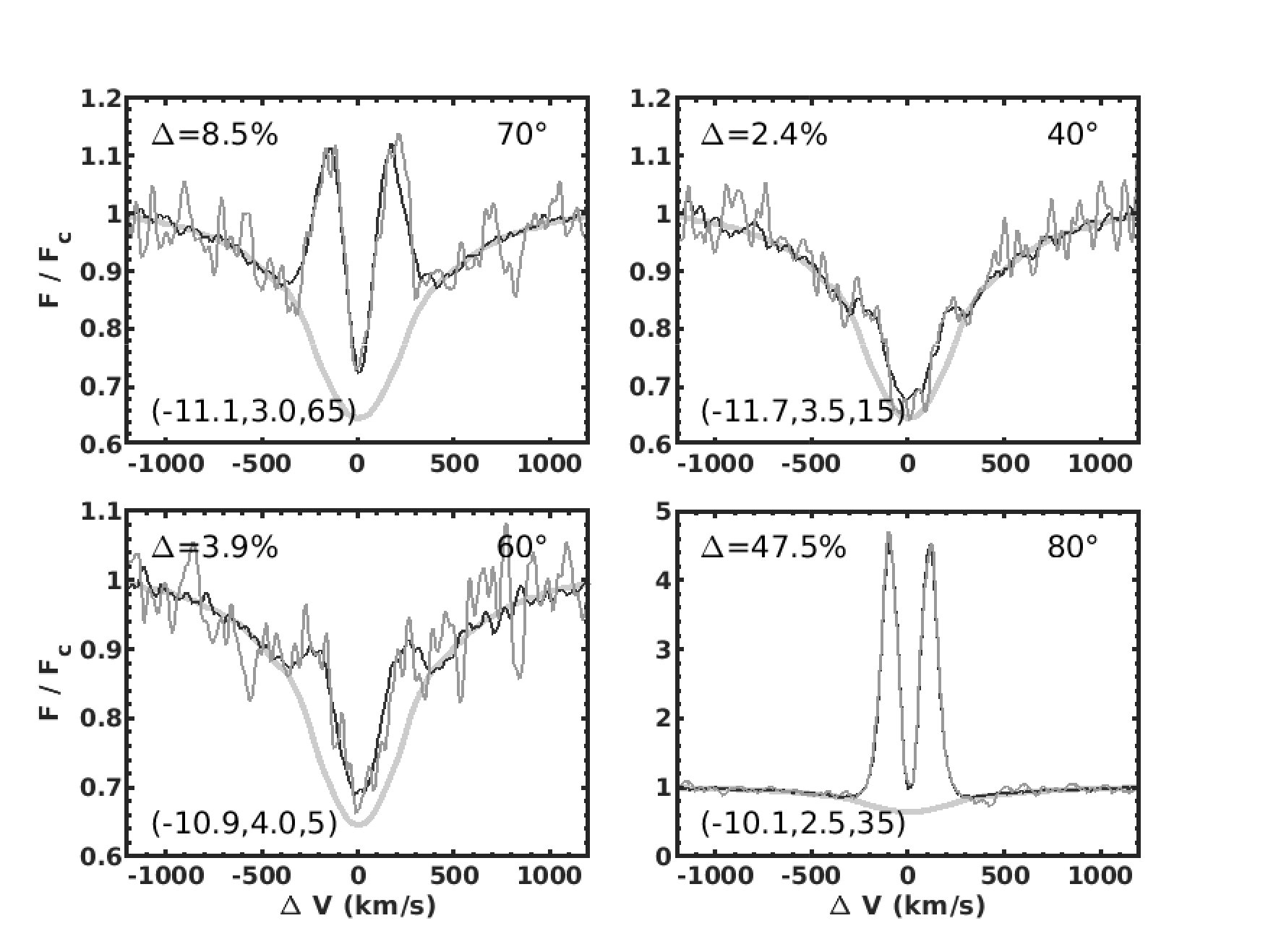}
\caption{Example synthetic \ha\ line profiles computed with \texttt{Beray} for a 4$\rm \,M_\odot$ Be star. In each panel is shown the H$\alpha$ line profile for a resolution of $10^4$ for a S/N of 100 (thin black line) and 25 (thin grey line). Also shown in each panel is the reference photospheric H$\alpha$ profile (thick grey line, same in each panel). In the top left of each figure is the average, absolute percentage difference ($\Delta$) between the $\rm S/N=100$ profile and the reference photospheric profile with the average taken over the region $\pm 1000\,\rm km\,s^{-1}$; this should be compared to the $3$ percent threshold to keep the profile in the sample. The different H$\alpha$ profiles are due to the different viewing inclinations (top right in each panel) and different disc parameters (bottom left, listed as $\log\rho_0$, $n$, $R_{\rm d}$).}
\label{sample_line_profile}
\end{figure}

Figure~\ref{sample_line_profile} shows several synthetic \ha\ emission line profiles for a 4 $\rm M_{\odot}$ Be star at $\mathcal{R} = 10,000$. Illustrated are a representative range of synthetic profiles for different choices of $\rm S/N$, disc density parameters, and viewing inclinations, with the the upper-right panel showing a profile rejected for being too close to the underlying photospheric profile and within the 3 percent tolerance. 

\subsection{Preprocessing the input spectra}
Each synthetic spectrum is stored in a 201-element vector containing the continuum-normalized, relative fluxes equally spaced in the interval $\pm 1,000\, \rm km\,s^{-1}$ about line centre in \ha. These vectors of relative fluxes are used as input for both types of neural networks, regression and classification. However, unlike neural networks, support vector regression uses Euclidean distances (see Section \ref{svr_subsection}), and vector elements with relatively large values (such as profiles with large emission peaks) will dominate the distance calculations. For this reason, we have scaled each of the samples of $\sim$8,000 synthetic spectra such that all elements have a mean of zero and unit standard deviation prior to use as input for support vector regression.   

Each observed spectrum was visually centered on the vacuum value of \ha, $\lambda_0$. The wavelengths associated with each flux in a spectrum were converted to velocities using the Doppler formula relative to line centre, $v/c = \Delta\lambda/\lambda_0$, as the \ha\ line covers only a narrow range of wavelengths. Compared to retaining the full wavelength dependence, this simplification results in errors that are at most a tenth of the assumed spectral resolution (i.e., $3\,\rm km\,s^{-1}$ compared to $30\,\rm km\,s^{-1}$ for ${\cal R}=10^4$). The observed spectra were truncated to the range $\pm 1,000\, \rm km\,s^{-1}$ and the fluxes were interpolated so that each observed spectrum  lies on the same 201 point velocity grid as the synthetic spectra. As with the synthetic spectra, these vectors of relative fluxes are used, directly, as inputs for both types of neural networks but are standardised to zero mean and unit standard deviation before being used as input to support vector regression.       

\section{Algorithms and performance metrics}
\label{Algorithms}

This work uses three types of supervised machine learning algorithms to learn the relationship between \ha\  emission line profiles and $i$: neural networks tasked with regression, neural networks tasked with classification, and support vector regression. The algorithms are trained on grids of relative fluxes from synthetic Be star line profiles, in the vicinity of \ha, and the trained algorithms are then used to determine $i$ for observed Be stars. 

A performance metric is needed in order to quantify how well the relationship between \ha\ emission line profiles and $i$ has been learned. The performance metric used in this work is the root mean squared error (RMSE), defined as 
\begin{equation}
\mathrm{RMSE} = \left(\frac{1}{n}\sum_{j=1}^{n}(y_j - \hat{y}_j)^2\right)^{\frac{1}{2}},
\label{eqn:rmse}
\end{equation}
where $n$ is the number of Be star spectra in the sample, $\hat{y}$ are the inclination angle determinations of our machine learning algorithms, and $y$ are our target inclinations. Each spectrum in a sample has an associated target inclination known precisely from the \texttt{Beray} calculation. All sample spectra are uniformly distributed from $0^\circ$ to $90^\circ$ in steps of $10^\circ$. In Sections \ref{section:zorec} and \ref{section:6}, we calculate the RMSE performance of the machine learning algorithms on observed spectra. For observed spectra, the target inclinations, $y$, are the inclination angle determinations of another method (e.g., \ha\ profile fitting).  

\subsection{Neural networks}
\label{neural_networks}

A neural network (NN) is a supervised machine learning algorithm comprised of computational units called nodes organized in layers. In feed-forward configuration, every node is a linear combination of the nodes in the preceding layer followed by an application of a non-linear activation function $h$. A single layer NN receives the 201 relative \ha\ fluxes as an input vector, $\boldsymbol{x}$, and returns a scalar output variable, $h(\boldsymbol{x},\boldsymbol{w})$, via the equation   
\begin{equation}
\centering
h(\boldsymbol{x},\boldsymbol{w}) \equiv h(\sum_{j=0}^{N}w_j x_j)
\end{equation}
by finding $\boldsymbol{w}$, the vector of weights, that minimizes a loss function which quantifies the discrepancy between the target values and the output values determined by the NN during training \citep{bishop1995}. This formulation of the NN equation implicitly includes the bias, a constant offset term, as the element $w_0$ by defining $x_0 \equiv 1$. Information about the loss functions used in this work can be found in Section \ref{section:training}.    
Although regression is the natural task of a machine learning algorithm that outputs a continuous scalar such as $i$, this work uses both regression as well as classification NNs\footnote{The regression and classification NNs were implemented using \texttt{MATLAB} R2021a functions \texttt{train} and \texttt{patternnet}, respectively.}. The outputs of classifiers are not normally directly comparable to those of regressors. However, by choosing an activation function whose output has a probabilistic interpretation, a weighted average can be used to transform a classification NN's output to a continuous scalar, which can then be compared with the output of the regression algorithms using the same performance metric. Although a full discussion is beyond the scope of this work, the authors are aware that the validity of this approach, which requires interpreting the output of the classification NNs as measures of model confidence, is contested \citep{Gal2016, xing2019}. 

The NNs tasked with regression use the hyperbolic tangent function\footnote{The hyperbolic tangent activation function can experience a problem known as vanishing gradients, particularly in NNs with many hidden layers \citep{Hochreiter1998}. We compared our NNs against otherwise identical NNs using the ReLU, $h(\sbullet) = \max(0,\sbullet)$, and leaky ReLU, $h(\sbullet) = \max(0.01,\sbullet)$, activation functions to ensure that vanishing gradients were not occurring.}   
\begin{equation}
\label{tanh}
h(\sbullet)=\frac{e^{\sbullet}-e^{-\sbullet}}{e^{\sbullet}+e^{-\sbullet}},
\end{equation}
as the activation function for each of their layers. In the above formulation, $\sbullet$ represents an arbitrary input. The NNs tasked with classification use two different activation functions. All of the layers other than the output layer use the hyperbolic tangent function, while the output layer uses the softmax function,
\begin{equation}
\centering
h(\sbullet) = \frac{e^{\sbullet}}{\sum e^{\sbullet }},
\end{equation}
where $\sbullet$ again represents an arbitrary input. The sum in the denominator is taken over the classes (which are the inclination bins $0^\circ$ to $90^\circ$ in steps of $10^\circ$ in this case) such that the denominator is a normalizing factor. This activation function was chosen because it assigns a probability to the likelihood that a given spectrum corresponds to each of the inclination classes. While regression NNs are the natural choice for this work (because $i$ is a continuous scalar), our primary reason for also including classification NNs is exploratory: we are interested in whether the softmax function outputs would be tightly clustered around the bins nearest to the target inclination or more flatly distributed.

\subsection{Support Vector Regression}
\label{svr_subsection}

Support vector regression (SVR) is a supervised machine learning algorithm that works by fitting a hyper-plane, with as many dimensions as the dataset contains features, to the data points. The SVR algorithm uses only a subset of the training data; data points sufficiently close to the hyper-plane (within a hyper-cylinder of radius $\boldsymbol{\varepsilon}$) are ignored \citep{Vapnik1999}. SVR was chosen for this work because it is deterministic, faster to train than NNs, and effective in high dimensional feature-spaces\footnote{SVR was implemented using the \texttt{MATLAB} R2021a function \texttt{fitrsvm}.}. 

SVR seeks to minimize
\begin{equation}
\centering
\frac{1}{2} || \boldsymbol{w} ||^2 +C\sum_{j=1}^{N}  (\zeta_j + \zeta_j^*)
\label{svr_eqn}
\end{equation}
with respect to $|| \boldsymbol{w} ||^2 $, subject to constraints
\begin{equation}
\centering
\begin{split}
y - \hat{y} \leq \boldsymbol{\varepsilon} + \zeta_j \\
\hat{y} - y \leq \boldsymbol{\varepsilon} + \zeta_j^* \\
\zeta_j, \zeta_j^* \geq 0,
\end{split}
\end{equation}
where $||\boldsymbol{w}||$ is the Euclidean norm of the vector of weights. Here $C$ is a regularization parameter, $y$ are the target values of $i$, $\hat{y}$ are the values of $i$ predicted by the model, and $\zeta^*$ and $\zeta$ are distances beginning at the border of the $\boldsymbol{\varepsilon}$-insensitive region and extending above and below it respectively \citep{Vapnik1999}.

\subsection{Committees of neural networks}

NNs initialized with random weights and biases can become trapped in poor local minima during training \citep{bishop2006}. Different NNs trained on the same inputs will, in general, have variance associated with their outputs even if the NNs are identically constructed \citep{bishop1995}. Furthermore, the RMSE is somewhat sensitive to outliers. To address these concerns, we train committees of independent NNs and retain only the median performing member. Two committees of five neural networks were trained for every central stellar mass listed in Table \ref{stellar_properties}; one committee is comprised of NNs tasked with regression and the other, tasked with classification. All 10 of the NNs associated with each central stellar mass are trained on the same sample of synthetic spectra (see Section \ref{samples_of_spectra} for details). 

Our approach differs from the commonly employed technique of bootstrap aggregation, whereby each neural network in the committee is trained on a bootstrapped sample of the original training sample and the overall determination of the committee is the average determination of its constituent members \citep{Breiman1996}. The advantage of bootstrap aggregation is that under ideal conditions (the errors of the committee members are uncorrelated and have a mean of zero), the average error of a committee falls like the reciprocal of the number of its constituent members \citep{bishop2006}. Unfortunately, these idealized conditions are not met in this work; the errors of the NNs are highly correlated and do not have a mean of zero (see Sections \ref{section:zorec} and \ref{section:6}), and we have instead chosen a committee structure that prioritizes outlier removal.     

\section{Hyper-Parameter Optimization}
\label{section:4}

The performance of machine learning algorithms on a given task varies depending on user-defined hyper-parameter values. Since the optimal values of these hyper-parameters are difficult to guess \textit{a priori} and can significantly impact performance, they must be searched for \citep{bishop2006}. The NN hyper-parameters that were optimized are the number of hidden layers ($n_l$) and the number of nodes per layer ($n_n$). The SVR hyper-parameters that were optimized are the size of the $\varepsilon$-insensitive region ($\varepsilon$), the regularization constant ($C$), and the kernel scale ($KS$). Additionally, all three algorithms have hyper-parameters that were chosen without being explicitly optimized in order to save computation time. These hyper-parameters were assigned standard choices and can be found in Section \ref{section:training}. 

The hyper-parameters of the machine learning algorithms were optimized independently for each of the Be star masses in Table \ref{stellar_properties}. Each Be star mass has an associated sample of 8,000 synthetic $\rm H\,\alpha$ profiles of $\mathcal{R} = 10,000$ and $\rm S/N = 25$. As there are 11 samples of synthetic profiles and three machine learning algorithms, this amounts to 33 sets of hyper-parameters to be optimized in total. 

Although the goal of this work is to produce an automated means of determining $i$ for observed Be stars from a single, medium-to-high resolution spectrum, the nature of the training process dictates that both the hyper-parameters and the parameters of the algorithms are optimized based on their ability to determine $i$ for synthetic spectra. The performances reported in this and the following section should be seen in that context.              
\subsection{Hyper-parameter optimization for NNs}
\label{h-pNN}

The hyper-parameters that were optimized for both types of NNs are the number of nodes per layer ($n_n$) and the number of hidden layers ($n_l$). For NNs tasked with regression, the optimization scheme consists of searching over a grid, found from preliminary trials, that contains the following six values of $n_n$, $n_n \in \{4, 5, 6, 8, 10, 12\}$, and two values of $n_l$, $n_l \in \{1, 2\}$. 

\begin{figure}
\centering
\includegraphics[width=0.47\textwidth]{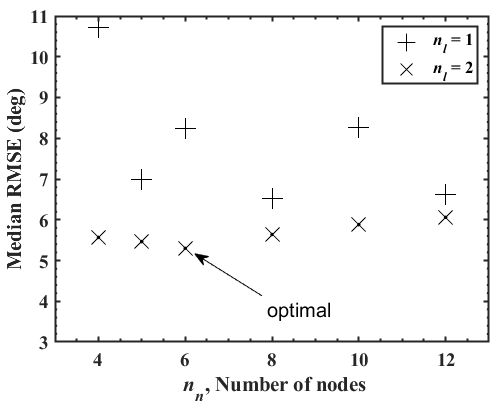}
\caption{RMSE performance versus number of nodes per hidden layer plot used to optimize the hyper-parameters for the 4 $\rm M_\odot$ sample for single (+) and dual (X) layer neural networks tasked with regression. Each data point represents the median performing member of a committee of five NNs. Hyper-parameters of two hidden layers of six nodes each were found to be optimal for this sample by virtue of having the lowest RMSE.}
\label{rmse_vs_nodes_plot}
\end{figure}

To perform the search, a committee of five NNs were trained on the same sample for each combination of $n_n$ and $n_l$ on the grid. The performance of a given ($n_n$,\,$n_l$) pair is taken to be the RMSE of its median performing committee member. The optimal hyper-parameters are taken to be the ($n_n$,\,$n_l$) pair with the best performance. Figure \ref{rmse_vs_nodes_plot} shows the hyper-parameter optimization scheme applied to the 4 $\rm M_\odot$ sample; here, the combination of two hidden layers of six nodes was found to be optimal. Then, this process was repeated for each of the remaining ten samples of synthetic profiles.

%\FloatBarrier

For NNs tasked with classification, the optimization scheme is nearly identical to that of the NNs tasked with regression. The only differences are that preliminary trials found that the grid to be searched over contains the following seven values of $n_n$, $n_n \in \{25, 30, 35, 40, 45, 50, 55\}$, and that the output of the classifier is a vector of probabilities that needs to be converted to an estimate of $i$ using a weighted average before a performance can be assigned via the RMSE. The optimal hyper-parameter combinations for both types of NNs are summarized in Table \ref{nn_optimisation_table}. While the performance always increased going from one to two hidden layers, we chose to limit the NN depth to two hidden layers because preliminary testing found that adding a third hidden layer rendered computation times prohibitive for only minimal gains in performance.     

\subsection{Hyper-parameter optimization for SVR}
\label{h-pSVR}

The hyper-parameters that were optimized for SVR are the $\varepsilon$-insensitive region ($\varepsilon$), the regularization constant of Equation (\ref{svr_eqn}) ($C$), and a scaling factor that the input matrix is divided by called the kernel scale (KS). For SVR, the optimization scheme consists of searching over combinations of $\varepsilon$, $C$, and $KS$ that were drawn randomly in log-space from the ranges 
\begin{equation}
\varepsilon \in [10^{-\frac{1}{2}}, 10^{\frac{1}{2}}] \frac{3\bar{y}}{S/N}\sqrt{\frac{\ln{N}}{N}},  
\end{equation}
\begin{equation}
C \in [10^{-\frac{1}{2}}, 10^{\frac{1}{2}}]|\bar{y} + 3\sigma_y|,  
\end{equation}
\begin{equation}
KS \in [15, 25], 
\end{equation}
where $\varepsilon$ and $C$ are drawn from a range spanning an order of magnitude from the prescription of \cite{CHERKASSKY2004113} and the range for $KS$ was determined from empirical trials.  

A combination of hyper-parameters is generated by drawing each of the three hyper-parameters independently. Once a combination of hyper-parameters has been drawn, an SVR is trained on one of the samples of synthetic profiles and its performance is stored. This process is repeated 150 times for each of the 11 samples of synthetic profiles. The motivation for repeating the process $n=150$ times is that it will find a hyper-parameter combination in the ninety-eighth percentile with $95$ percent confidence via solving $1-0.98^n = 0.95$ for $n\approx 148$. We determined that significant performance gains were unlikely to be achieved by raising the value of $n$ by comparing the three best hyper-parameter combinations for each of the 11 samples and noting that the performance difference between the best and third best hyper-parameter combination was always below $0.1^\circ$. The optimal SVR hyper-parameter combinations, resulting from this process, are summarized in Table \ref{svr_optimisation_table}.   

\section{Training the Algorithms}
\label{section:training}

This section describes how the machine learning algorithms are trained on samples of synthetic spectra. By modifying their adaptive parameters, namely, weights and biases, training allows the machine learning algorithms to leverage patterns between the synthetic $\rm H\,\alpha$ profiles and their associated inclination angles. The trained algorithms will then be used to determined the inclination angles of observed Be stars in the following two sections. 

The training process introduces additional hyper-parameters to those optimized in Section \ref{section:4}. These hyper-parameters, which have been assigned standard choices, are the loss function, training algorithm, and kernel function. While the work is organized such that Section \ref{section:4} is about hyper-parameter optimization and Section \ref{section:training} is about training the algorithms, the two sections should be seen as complimentary: the optimized hyper-parameters are used during training and the training process was used to optimize the hyper-parameters.   

\subsection{loss functions}
\label{error_functions}

In order to quantify the discrepancy between the determinations of a machine learning algorithm and their associated target inclinations during training, a loss function, $E(\boldsymbol{w})$, is used. For a machine learning algorithm, learning consists of minimizing the loss function by modifying the adaptive parameters of the algorithm, namely the weights and biases. 

For both NNs tasked with regression and SVR, this work uses the mean squared error, 
\begin{equation}
E(\boldsymbol{w}) = \frac{1}{n}\sum_{j=1}^{n}(y_j - \hat{y}_j)^2,
\label{eqn:mse}
\end{equation}
where $n$ is the number of Be star spectra in the sample, $\hat{y}$ are the inclination angle determinations of our models, and $y$ are the target inclinations, as the loss function. The mean squared error was chosen because it contains the same information as our performance metric, is a standard choice for regression problems, and because it has the property of heavily penalizing large errors.  

For NNs tasked with classification, this work uses cross-entropy,
\begin{equation}
E(\boldsymbol{w}) = -\frac{1}{n}\sum_{j=1}^{n}\sum_{k}p(y_{jk})\ln({p(\hat{y}_{jk}})) + (1-p(y_{jk}))\ln({1-p(\hat{y}_{jk}})) ,
\label{eqn:ce}
\end{equation}
where $n$, $\hat{y}$, and $y$ are defined as in equation (\ref{eqn:mse}) and the second sum is taken over the classes. The probability that a profile's target inclination belongs to a given class, $p(y)$, can only have values of zero or one. If we consider a profile with an associated target inclination of $y = 20^\circ$, then $p(y)$ is equal to one when $k$ corresponds to the $20^\circ$ class and is equal to zero otherwise. The inclination determinations of the classifiers are vectors of length $k$, whose components contain the probability that a profile belongs to each of the inclination classes, $p(\hat{y})$. While cross-entropy is the standard loss function used in classification NNs, recent work has cast doubt on its, supposed, superiority over the mean squared error \citep{Muthukumar2021,hui2021}. Nevertheless, cross-entropy was chosen because using a squared loss function appears to impede the optimization of NNs with a softmax output layer \citep{hui2021} which is required for converting the probability that each profile belongs to a given inclination class into a scalar estimation of $i$ (see Section \ref{neural_networks}).                

\subsection{Optimization algorithms}

In order to minimize the loss functions discussed in Section \ref{error_functions}, an optimization algorithm is required. The two major classes of algorithms applicable to minimising continuous, differentiable functions of several variables (applicable to both types of NNs) are variants of either gradient descent or Newton's method. The optimization algorithm we use for the NNs tasked with regression, which is effectively an interpolation between these two major classes of algorithms, is the \texttt{MATLAB} R2021a implementation of the Levenberg--Marquardt algorithm \citep{Levenberg1944,marquardt:1963}. 

In order to reduce computational cost, the Levenberg--Marquardt algorithm uses $\mathbf{J^TJ}$, where $\mathbf{J}$ is the Jacobian matrix, to approximate the Hessian matrix when performing Newton's method-like parameter updates; this approximation only holds for squared loss functions and is, therefore, incompatible with the NNs tasked with classification (which use cross-entropy as their loss function). The optimization algorithm we use for the NNs tasked with classification, which is an accelerated variant of gradient descent, is the \texttt{MATLAB} R2021a implementation of the scaled conjugate gradient descent algorithm \citep{Moller1993}. 

SVR results in a very large convex, quadratic programming (QP) optimization problem. As the optimization surface is convex, SVR cannot become trapped in poor local minima during training the way NNs can. The optimization algorithm we use for SVR, which breaks this large QP problem into a series of minimally sized QP problems that can be solved analytically, is the \texttt{MATLAB} R2021a implementation of sequential minimal optimization \citep{Platt1998}\footnote{The Levenberg--Marquardt, scaled conjugate gradient descent, and sequential minimal optimization algorithms were implemented using the \texttt{MATLAB} R2021a functions \texttt{trainlm}, \texttt{trainscg}, and \texttt{smo}, respectively.}. 

\subsection{Kernel function}

For SVR, the kernel function, $K$, maps inputs to higher dimensional spaces, where a suitable hyper-plane can be found, before back-projecting to the original feature space \citep{bishop2006}. This allows for a curved hyper-plane, which may provide a significantly better fit to the data than a straight one would. The radial basis function
\begin{equation}
    K(\boldsymbol{x},\boldsymbol{x'}) = e^{-||\boldsymbol{x}-\boldsymbol{x'}||^2},
\end{equation}
where $\boldsymbol{x}$ and $\boldsymbol{x'}$ represent feature vectors of relative fluxes, is the kernel function used in this work. The radial basis function was chosen because it is the standard kernel function used in SVR and has good performance over a wide variety of tasks \citep{Liu2014}. 

\begin{table}
\centering
\begin{tabular}{ccccccc} 
\hline
 Mass & \multicolumn{3}{c}{\hrulefill~Regression~\hrulefill} & \multicolumn{3}{c}{\hrulefill~Classification~\hrulefill} \\
$(\rm M_\odot)$  & $n_n$ & $n_l$ & RSME$(^\circ)$ & $n_n$ & $n_l$ & RMSE$(^\circ)$  \\
\hline
3 & 6 & 2 & 3.7 & 45 & 2  & 6.7\\ 
4 & 6 & 2 & 5.3 & 45 & 2  & 6.7 \\   
5 & 6 & 2 & 6.1 & 40 & 2  & 7.7\\  
6 & 6 & 2 & 6.9 & 50 & 2  & 8.3\\  
7 & 8 & 2 & 7.0 & 50 & 2  & 8.6\\ 
8 & 5 & 2 & 8.1 & 30 & 2  & 9.3 \\ 
9 & 4 & 2 & 8.2 & 50 & 2  & 10.6\\ 
10 & 5 & 2 & 8.3 & 45 & 2 & 11.0 \\ 
12 & 5 & 2 & 8.4 & 45 & 2 & 11.4\\ 
14 & 4 & 2 & 8.1 & 40 & 2 & 11.3\\ 
16 & 4 & 2 & 8.2 & 40 & 2 & 11.2\\ \hline
\end{tabular}
\caption{The RMSE performance of regression and classification neural networks trained on $\mathcal{R} = 10,000$ and $\rm S/N = 25$  synthetic line profiles, for each central B star mass considered. The optimal hyper-parameters determined in Section \ref{h-pNN}, the number of nodes per layer ($n_n$) and the number of hidden layers ($n_l$), are also shown. 
}
\label{nn_optimisation_table}
\end{table}   

\begin{table}
\centering
\begin{tabular}{c c c c  c} 
\hline
Mass ($\rm M_\odot$) & $\epsilon$ & $C$ & $KS$ & RMSE ($^\circ$) \\ [0.47ex] 
\hline
3 & 0.567 & 220 & 19.2 & 6.0 \\ 
4 &  0.522 & 177 & 17.8 & 6.0 \\   
5 &  0.561 & 131 & 16.7 & 6.6\\  
6 &  0.546 & 122 & 16.8 & 7.3 \\  
7 &  0.572 & 104 & 16.6 & 7.5 \\ 
8 &  0.557 & 121 & 16.2 & 7.5 \\ 
9 &  0.566 & 97.5 & 15.4 & 7.8 \\ 
10 &  0.566 & 106 & 16.0 & 8.1 \\ 
12 &  0.563 & 92.2 & 15.5 & 8.3 \\ 
14 &  0.564 & 108 & 15.1 & 8.1 \\ 
16 &  0.569 & 100 & 15.8 & 8.2 \\ \hline
\end{tabular}
\caption{The RMSE performance of SVR trained on $\mathcal{R} = 10,000$ and $\rm S/N = 25$  synthetic line profiles, for each central B star mass considered. The optimal hyper-parameters determined in Section \ref{h-pSVR}, the epsilon insensitive region ($\epsilon$), the regularization constant (C), and the kernel scale (KS), are also shown. 
}
\label{svr_optimisation_table}
\end{table}

\subsection{Training, validation, and testing results}
\label{train,val,test}

Following standard practice in machine learning, we divided each of the samples of synthetic spectra (see Section \ref{samples_of_spectra}) into disjoint training, validation, and testing datasets \citep{Hastie2009}:

\begin{enumerate}

    \item The largest of the three datasets is the training set. For both types of NNs, we randomly assigned $70$ percent of each sample of synthetic spectra to the training set; for SVR, this percentage is higher, at $90$ percent, owing to the different validation methods used. For a machine learning algorithm, training consists of using the profiles of the training set to learn the model parameters that minimize $E(\boldsymbol{w})$. The performances that the algorithms achieve on the training set are prone to being exceedingly optimistic due to over-fitting. Over-fitting occurs when an algorithm becomes over-specialized to the peculiarities of the training data (such as noise) and, as a result, generalizes poorly to new data. 
    
    \item The validation set is held back during training and is used to prevent over-fitting rather than to modify the adaptive parameters of the algorithms. For both types of NNs, we randomly assigned $15$ percent of each sample of synthetic spectra to the validation set. Validation is performed by calculating $E(\boldsymbol{w})$ on the validation set each time the model parameters are updated during training. If a NN is over-fitting, $E(\boldsymbol{w})$ will decrease on the training set but increase on the validation set due to poor generalization to new data. If $E(\boldsymbol{w})$ increases for six consecutive parameter updates on the validation set, the model is considered over-fitted and the training ends via a validation criterion known as early stopping \citep{bishop2006}. The number of parameter-updates performed during training before over-fitting began is stored as the `best epoch' parameter for later use. 
    
    For SVR, we used ten-fold cross validation whereby the training set is randomly divided into ten equally sized subsets called folds \citep{Hastie2009}. The SVR model is trained ten different times; each of these times a different fold is held back to be used as a validation set and the remaining nine folds are combined into a training set. Validation is performed by calculating $E(\boldsymbol{w})$, averaged over the ten validation sets, each time the model parameters are updated during training. The number of parameter-updates that minimizes $E(\boldsymbol{w})$ is stored as the `best epoch' parameter for later use. Ten-fold cross validation has the advantage that every profile in the training set contributes to both training and validation, but comes at the cost of significantly increased training times. This trade-off was ideal for SVR, which is relatively quick to train, but computationally prohibitive for the NNs.

    \item The test set is held back during both training and validation and is used to test the trained algorithms' performance on previously unseen data. Any profile that did not end up in either the training or validation set was assigned to the test set; this amounted to $15$ percent of each sample of synthetic spectra for both types of NNs and $10$ percent for SVR. Testing is performed by calculating the RMSE performance of an algorithm that was trained for a number of parameter updates defined by its associated `best epoch' parameter. 
    
\end{enumerate}
In this work, the performance of an algorithm on synthetic spectra (in Tables~\ref{nn_optimisation_table}, and \ref{svr_optimisation_table}, and in Figure \ref{rmse_vs_nodes_plot}) always refers to the test set performance. 

%\subsection{Test Sample Results}
%\label{section:5}

Tables~\ref{nn_optimisation_table} and \ref{svr_optimisation_table} summarize the RMSE performance of the three machine learning algorithms on the test samples of synthetic spectra for each mass bin. There is a trend that the RMSE of all three machine learning algorithms tends to worsen as stellar mass increases until it plateaus at around 9--10 $\rm M_\odot$. The two regression algorithms outperformed the NNs tasked with classification for every stellar mass considered. The NNs tasked with regression outperformed SVR for masses between three and seven $\rm M_\odot$ whereas SVR outperformed the NNs tasked with regression for eight and nine $\rm M_\odot$ spectra; the performance of the two regression algorithms is approximately equal at 10 $\rm M_\odot$ and above.   

\section{Performance on observed profiles}
\label{section:zorec}

The results of the previous section are an encouraging proof-of-concept. However, the method must still be shown to be effective on observational data. This section is concerned with testing the trained algorithms on an observed sample of \ha\ spectra consisting of 92 of the 233 galactic Be stars considered by \cite{Zorec2016}, which we call the Zorec sample {following \citet{Sigut2022}}. The stars of the Zorec sample were chosen based on the {public} availability of spectra in the region of \ha. Spectra for 58 of the stars come from the BeSS spectral database\footnote{Operated at LESIA, Observatoire de Meudon, France:} \url{http://basebe.obspm.fr}, with the remaining 34 stars coming from the sample of \citet{Silaj2010} taken at the John Hall telescope at Lowell Observatory. Sources for individual stars can be found in Table~1 of \citet{Sigut2022}. The spectra typically have $\rm S/N\sim100$ and $\mathcal{R}\sim10^4$, with the latter matching the training resolution. Every star in the sample has an inclination angle determination based on gravitational darkening \citep{Zorec2016} and \ha\ profile fitting \citep{Sigut2022}. More information on the 92 stars of the Zorec sample can be found by consulting \cite{Sigut2022} and the references therein. 

\begin{figure}
\centering
\includegraphics[width=0.47\textwidth]{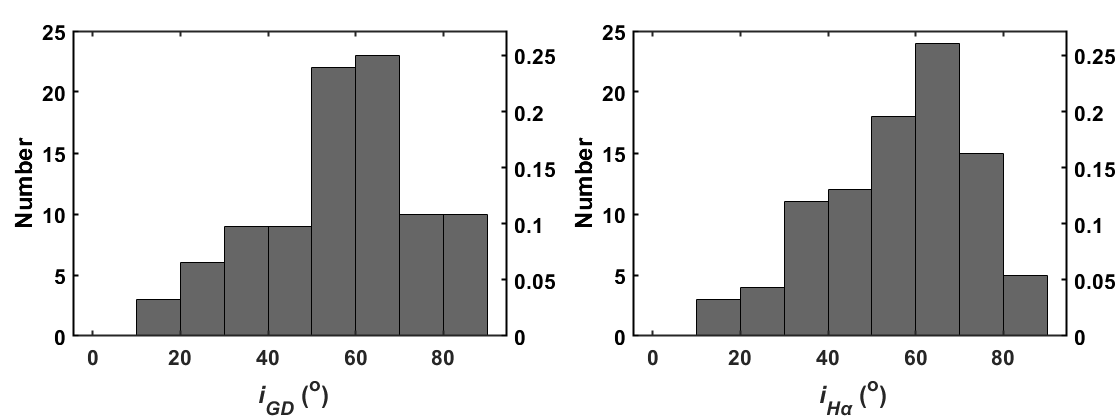}
\caption{Inclination angle histograms of the 92 Be stars of the Zorec sample as determined by gravity darkening (left) and \iha\ profile fitting (right). The x-axis shows inclinations in nine bins of $10^\circ$, the left y-axis shows the absolute number of stars, and the right y-axis shows the fractional number of stars. Note the different shapes of the distributions showing the trend of higher values of \iha\ as compared to \igd\ at low inclinations and vice versa.}
\label{zorec:hist}
\end{figure}

The \cite{Zorec2016} inclination angle determinations (referred to as \igd\ hereafter) are based on gravity darkening whereby very rapid stellar rotation results in a latitude-dependent $T_{\rm eff}$ \citep{vonZeipel1924}, causing the spectrum to vary with $i$. Figure~\ref{zorec:hist} shows the inclination angle distribution of the Zorec sample as determined by both gravity darkening (\igd, left) and \ha\ profile fitting (\iha, right). Although both distributions peak near $60^\circ$, there is a trend for \iha\ to be higher for low inclinations and lower for high inclinations than \igd. As a comparison between \iha\ and \igd\ on the stars of the Zorec sample has already been done \citet{Sigut2022}\footnote{\citet{Sigut2022} carefully discusses the apparent non-$\sin i$ distributions of both \igd\ and \iha.}, this section will focus on a comparison between the machine learning determinations of $i$ and \iha.

All three trained machine learning algorithms discussed in Section \ref{section:training} were used to determine the inclination angles of the Zorec sample stars. These inclination angles were calibrated and then compared with \iha\ to ascertain how effectively the different machine learning algorithms, trained on only synthetic spectra, can determine inclinations using observed spectra. To calibrate the machine learning inclinations, the mean of the distribution ($i_{\rm ML}-i_{\rm H\,\alpha}$) was set to zero by adding a constant offset to $i_{\rm ML}$. Here $i_{\rm ML}$ refers to inclinations determined using each of the three algorithms, NNs tasked with regression, NNs tasked with classification, and SVR. These calibration offsets are given in Table~\ref{zorec_calibration_offsets} for each of the three machine learning algorithms. We note that two of these offsets, NNs tasked with regression and SVR, are quite small. NNs tasked with classification has the largest offset of $-7.3^\circ$; however, even in this case, the offset is still less than most of the $1\sigma$ errors in \iha\ as determined by \citet{Sigut2022}.     

\begin{table}
\centering
\begin{tabular}{c c c } 
\hline
  \multicolumn{3}{|c|}{Additive Calibration Offsets} \\ [0.47ex] 
 $i_{\rm NN}$ ($^\circ$) & $i_{\rm CNN}$ ($^\circ$) & $i_{\rm SVR}$ ($^\circ$) \\
\hline
-3.4 & -7.3 & +0.4  \\ 
\hline
\end{tabular}
\caption{Calibration offsets applied to the three different machine learning algorithms. These offsets were determined by forcing the mean of the distribution $i_{\rm ML}-i_{\rm H\,\alpha}$ to zero for the stars of the Zorec sample.}
\label{zorec_calibration_offsets}
\end{table}

%\subsection{Comparisons with \texorpdfstring{H$\alpha$}{Ha} profile fitting}
\label{zorec_comparison_iHA}

\begin{figure*}
\centering
\includegraphics[width=0.96\textwidth ]{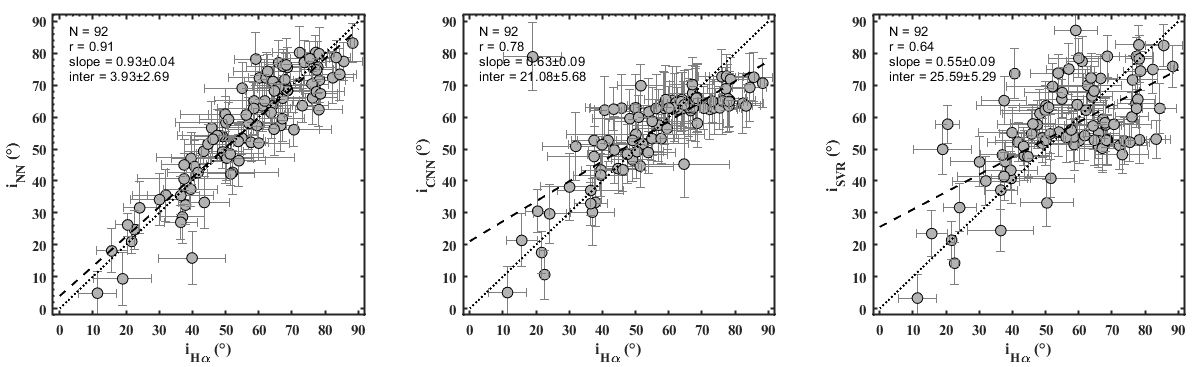}
\caption{Inclinations (in degrees) $i_{NN}$ (left), $i_{CNN}$ (middle), and $i_{SVR}$ (right) versus $i_{\rm H\,\alpha}$ for the 92 stars of the Zorec sample. The horizontal error bars show the $1\sigma$ uncertainties determined by \citet{Sigut2022} and the vertical error bars show the test set RMSEs (see Tables \ref{nn_optimisation_table} and \ref{svr_optimisation_table}). The dashed lines show the least squares fits to the data and the dotted lines show a slope of one for comparison. The Pearson correlation coefficients, as well as the slopes and intercepts of the least squares fits, can be found in the upper-left area of each plot.}
\label{zorec_halpha_panel_plot}
\end{figure*}

\begin{figure}
\centering
\includegraphics[width=0.45\textwidth]{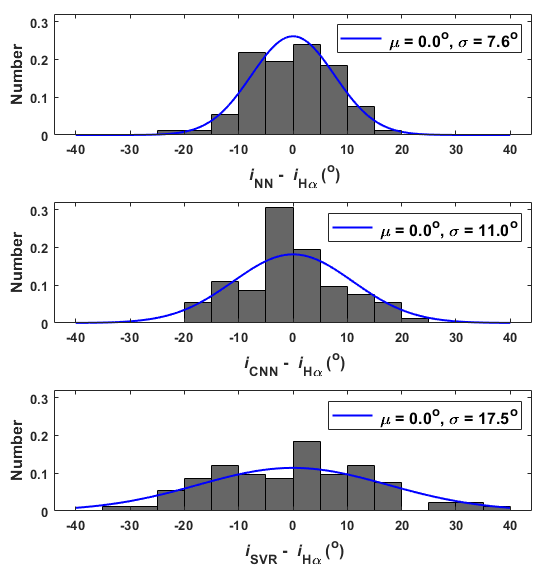}
\caption{Residual histograms of $i_{\rm NN} - i_{\rm H\,\alpha}$ (top), $i_{\rm CNN} - i_{\rm H\,\alpha}$ (middle), and $i_{\rm SVR} - i_{\rm H\,\alpha}$ (bottom) for the Zorec sample. The x-axis shows residuals in bins of $5^\circ$ and the y-axis shows the fractional number of stars per bin. The blue curve is a Gaussian with the same mean and standard deviation as the distribution of the residuals (shown upper right). Note that the means were calibrated to be $0.0^\circ$ (Table \ref{zorec_calibration_offsets}) and one star, HD58050, is missing from the middle panel to improve readability as $i_{\rm CNN} - i_{\rm H\,\alpha}$ was very large at $+63.1^\circ$.}
\label{zorec_hist_gauss}
\end{figure}

Figure~\ref{zorec_halpha_panel_plot} plots the inclination angle determinations of the three types of algorithms: NNs tasked with regression ($i_{\rm NN}$), NNs tasked with classification ($i_{\rm CNN}$), and SVR ($i_{\rm SVR}$), each versus the corresponding \iha\ for each of the 92 stars of the Zorec sample. The $1\sigma$ uncertainties in \iha\ are as determined by \cite{Sigut2022}. We have adopted the algorithms' RMSE performance on synthetic spectra of an equivalent mass star (see Tables~\ref{nn_optimisation_table} and \ref{svr_optimisation_table}) as their $1\sigma$ uncertainties. The Pearson correlation coefficients, $r$, were calculated for each of the three plots, as were least-squares fits to the data, including uncertainties from bootstrap Monte Carlo resampling done 100 times. Figure~\ref{zorec_hist_gauss} shows the distribution of the residuals between the inclinations determined by each algorithm and \iha\ for the stars of the Zorec sample (i.e, $i_{\rm NN} - i_{\rm H\,\alpha}$, $i_{\rm CNN} - i_{\rm H\,\alpha}$, and $i_{\rm SVR} - i_{\rm H\,\alpha}$). Theses residuals are binned in widths of $5^\circ$. The blue curve in each plot shows a Gaussian distribution with the same mean and standard deviation as the distribution of the residuals for comparison. 

%Figure~\ref{zorec_hist_gauss} shows the distribution of the residuals between the inclinations determined by each algorithm and \iha\ for the stars of the Zorec sample (i.e, $i_{\rm NN} - i_{\rm H\,\alpha}$, $i_{\rm CNN} - i_{\rm H\,\alpha}$, and $i_{\rm SVR} - i_{\rm H\,\alpha}$). Theses residuals are binned in widths of $5^\circ$. The blue curve in each plot shows a Gaussian distribution with the same mean and standard deviation as the distribution of the residuals for comparison. 

Figures~\ref{zorec_halpha_panel_plot} and \ref{zorec_hist_gauss} show a clear hierarchy of performance; the NNs tasked with regression outperformed the NNs tasked with classification which in turn outperformed SVR. The NNs tasked with regression performed the best with a RMSE of $7.6^\circ$ and a correlation coefficient between $i_{\rm NN}$ and \iha\ of $r=+0.91$. Of the 92 stars, 78 (or 85 percent) were found to have $(i_{\rm NN}-i_{\rm H\,\alpha})$ consistent with zero within the errors. The NNs tasked with classification had an intermediary performance with a RMSE of $10.9^\circ$ and a correlation coefficient of $r=+0.78$; 71 of the 92 stars (or 77 percent) were found to have $(i_{\rm CNN}-i_{\rm H\,\alpha})$ consistent with zero within the errors. Finally, SVR performed notably worse than the NNs. with a RMSE of $13.9^\circ$. The correlation coefficient was found to be $r=+0.64$, and 47 of the 92 stars (or 51 percent) were found to have $(i_{\rm SVR}-i_{\rm H\,\alpha})$ consistent with zero within the errors. Thus NNs tasked with regression are the optimal choice, providing an accuracy comparable to the direct \ha\ profile fitting of \citet{Sigut2022}. 

\subsection{Performance by mass and inclination}
\label{zorec:mass_inc}

Overall, NNs tasked with regression best automates the method of \iha\ profile fitting; however, it is possible that one of the other algorithms is better at determining $i$ for Be stars with particular properties. Should this be the case, the best approach would not rely on a single ``best" algorithm but would instead be an ensemble of two (or all three) algorithms whose outputs would be weighted by the properties of the star of interest. This subsection will look with more granularity at two such properties, stellar mass and inclination, with the goal of determining if either the NNs tasked with classification or SVR can outperform the NNs tasked with regression on particular mass and/or inclination ranges. 

We have designated the stars of the Zorec sample as either low mass (3--5$\,M_\odot$, N=36), medium mass (6--8$\,M_\odot$, N=24), or high mass (9--14$\,M_\odot$, N=32) and tabulated\footnote{We use the mass-spectral type calibration of \citet{Sigut2022}.} the algorithms' performances in Table~\ref{zorec_mass_table}. When tested on synthetic spectra (Section \ref{section:training}), there was a trend that the performance of all three algorithms tended to worsen as stellar mass increased until it plateaued around 9--10$\,M_\odot$ (see Tables~\ref{nn_optimisation_table} and \ref{svr_optimisation_table}). With observed spectra, NNs tasked with regression performed similarly on both the observed spectra of low ($\rm RMSE = 7.0^\circ$) and medium mass ($\rm RMSE = 6.7^\circ$) stars, with performance worsening for the high mass stars ($\rm RMSE = 8.8^\circ$). NNs tasked with classification performed worse on the observed spectra of low mass stars ($\rm RMSE = 9.1^\circ$) compared to medium masses ($\rm RMSE = 8.0^\circ$), with their performance worsening for high mass stars ($\rm RMSE = 14.2^\circ$). SVR performed best on the observed spectra of low mass stars ($\rm RMSE = 13.4^\circ$) and performed similarly on both medium ($\rm RMSE = 14.2^\circ$) and high mass stars ($\rm RMSE = 14.3^\circ$). Ultimately, however, the NNs tasked with regression outperformed both of the other algorithms on all three mass ranges suggesting that an ensemble of the algorithms is not warranted based on mass.   

\begin{table}
\centering
\begin{tabular}{c c c c } 
\hline
& Low Mass & Medium Mass  & High Mass   \\ [0.47ex] 
& (3--5 $\rm M_\odot$) & (6--8 $\rm M_\odot$) & (9--14 $\rm M_\odot$) \\
& RMSE ($^\circ$) & RMSE ($^\circ$) & RMSE ($^\circ$) \\
\hline
NN& 7.0 & 6.7 & 8.8  \\ 
CNN& 9.1 &  8.0 & 14.2  \\      
SVR & 13.4 &  14.2 & 14.3 \\  
\hline
\end{tabular}
\caption{RMSE performance, in degrees, of the three algorithms on the 92 observed Be stars of the Zorec sample subdivided by mass into low, medium, and high mass stars.}
\label{zorec_mass_table}
\end{table}

Turning now to inclination, we have designated the stars of the Zorec sample as either low~$i$ (0--30$^\circ$, N=7), medium~$i$ (30--60$^\circ$, N=41), or high~$i$ (60--90$^\circ$, N=44) and tabulated the three algorithms' performances in Table~\ref{zorec_inc_table}. The small sample size of low inclination stars is unfortunate but not surprising because of the $p(i)\sim\sin{i}$ for randomly oriented spin axes \citep{gray2021}. The NNs tasked with regression performed best on low $i$ observed spectra ($\rm RMSE = 5.8^\circ$) and similarly on both medium ($\rm RMSE = 8.0^\circ$) and high $i$ stars ($\rm RMSE = 7.5^\circ$). The NNs tasked with classification performed the worst on low $i$ ($\rm RMSE = 23.8^\circ$) observed spectra and similarly on both medium ($\rm RMSE = 9.2^\circ$) and high $i$ stars ($\rm RMSE = 9.0^\circ$). The very poor performance of the NNs tasked with classification on low $i$ observed spectra is the result of a small sample size ($N=7$) combined with the worst determination of $i$ of any algorithm on the Zorec sample for the star HD\,58050 (with $i_{\rm CNN} - i_{\rm H\,\alpha} = +63.1^\circ$); omitting HD\,58050 improves the performance considerably ($\rm RMSE = 7.8^\circ$). SVR performed worse on low $i$ ($\rm RMSE = 19.3^\circ$) than on medium $i$ ($\rm RMSE = 12.7^\circ$) observed spectra with an intermediate performance for high mass stars ($\rm RMSE = 14.0^\circ$). The NNs tasked with regression outperformed both of the other algorithms on all three inclination ranges confirming that an ensemble of algorithms is not warranted for this task either. 

\begin{table}
\centering
\begin{tabular}{c c c c } 
\hline
& Low \textit{i} (0--30$^\circ$) & Medium \textit{i} (30--60$^\circ$)  & High \textit{i} (60--90$^\circ$)  \\ [0.47ex] 
& RMSE ($^\circ$) & RMSE ($^\circ$) & RMSE ($^\circ$) \\
\hline
NN& 5.8 & 8.0 & 7.5  \\ 
CNN& 23.8 (7.8) &  9.2 & 9.0  \\   
SVR & 19.3 &  12.7 & 14.0 \\  
\hline
\end{tabular}
\caption{RMSE performance, in degrees, of the three algorithms on the 92 observed Be stars of the Zorec sample subdivided by inclination angle into low, medium, and high inclinations. It is worth noting that for NNs tasked with classification, the combination of a small sample size ($\rm N = 7$) and a very poor determination of the star HD\,58050 ($i_{\rm CNN} - i_{\rm H\,\alpha} = +63.1^\circ$) has resulted in a very poor performance on low inclination stars which may be misleading; included in parentheses, is the performance with HD\,58050 omitted.}
\label{zorec_inc_table}
\end{table}

\subsection{Discussion}
\label{zorec:discussion}

While NNs tasked with regression and SVR performed similarly on synthetic spectra (Section \ref{section:training}), NNs tasked with regression performed significantly better than SVR at automating the results of \cite{Sigut2020}'s \iha\ profile fitting method on observed Be star spectra. It is also interesting to note that the NNs tasked with classification, the worst performer on synthetic spectra, actually outperformed SVR on observed Be star spectra. With an RMSE of $7.6^\circ$ and a Pearson coefficient of $r=+0.91$, the NNs tasked with regression are the clear choice to automate the \iha\ profile fitting method. The NNs tasked with regression's performance of $\rm RMSE = 7.6^\circ$ matches the average uncertainty in \iha\ ($\Delta i_{\rm H\,\alpha} = 7.6^\circ$) on the Zorec sample, again, suggesting excellent agreement between the two methods. An ensemble of specialists was considered in Section \ref{zorec:mass_inc}, but ultimately rejected because the NNs tasked with regression had the best performance on every subdivision of mass and inclination considered. 

\section{The NPOI Sample}
\label{section:6}

This section is concerned with testing the calibrated algorithms on an observational sample of 11 bright, nearby, Be stars taken by the Naval Precision Optical Interferometer (NPOI) \citep{Armstrong_1998} {with \ha\ spectra available from \citet{Silaj2010}}. The NPOI observations feature the spatially resolved circumstellar discs of their associated Be stars which allows for accurate determinations of their inclination angles. If $a$ is the measured major axis of the disc and $b$ is the minor axis, we can calculate the interferometrically determined inclination angle via $i_{\rm NPOI}=\cos^{-1}(b/a)$ on the simple geometric assumption that the disc is circular yet appears elliptical due to projection. While it is well established that Be star discs are thin \citep{Porter2003}, they do have a small associated scale height. Therefore, interferometric observations of sufficient angular resolution can never yield $b = 0$ and we should take care to only use $i_{\rm NPOI}$ for inclinations where it is appropriate. \cite{Sigut2020} examines when the $\cos^{-1}(b/a)$ relation is expected to fail and finds it to be when $i_{\rm NPOI} > 80^\circ$. None of the 11 Be stars in the NPOI sample have an inclination value outside this range, so it is used for all the determinations of $i_{\rm NPOI}$ in this work. More information about the 11 stars in the NPOI sample can be found by consulting Table \ref{NPOI_properties}.  

The main advantage of the NPOI sample is the high accuracy of the interferometrically-determined inclinations, which have average uncertainties that are about two and a half times smaller than the uncertainties of the inclinations determined by gravity darkening ($5.5^\circ$ vs $14.5^\circ$). Furthermore, unlike \iha\ profile fitting, the method of interferometry is entirely independent of the \ha\ spectroscopy used to train the algorithms used in this work. The main disadvantage of the NPOI sample is its small size; as only the brightest and closest Be stars can be resolved interferometrically, the resulting sample of 11 profiles will necessarily be sensitive to outliers. 

\begin{table}
\centering
\begin{tabular}{c c c c c c c} 
\hline
Name & HD & Sp Type & Mass & $i_{\rm NPOI}$ & $i_{\rm H\,\alpha}$ & $i_{\rm GD}$\\
 & & & ($\rm M_{\odot}$) &  ($^{\circ}$) & ($^{\circ}$) & ($^{\circ}$)
\\ [0.49ex] 
\hline
$\gamma$ Cas & 5394 & B0.5IV & 14.6 &  51.6$\pm$3.3 &  58.7$\pm$6.9 & 65$\pm16$\\ 
$\phi$ Per & 10516 & B1.5V & 11.0 & 74.0$\pm$0.6 & 69.1$\pm$6.0 & 57$\pm14$\\   
$\psi$ Per & 22192 & B5Ve & 5.5 & 71.2$\pm$0.5 &  72.5$\pm$4.6 & 74$\pm18$\\ 
$\eta$ Tau & 23630 & B7III & 4.2 & 33.0$\pm$3.2 &  50.0$\pm$5.0 & 62$\pm15$\\  
48 Per & 25940 & B3Ve & 7.6 & 45.0$\pm$3.1 &40.0$\pm$5.0 & 50$\pm12$\\  
$\beta$ CMi & 58715 & B8Ve & 3.8 & 46.0$\pm$9.1 & 45.0$\pm$5.3 & 66$\pm16$\\  
$\kappa$ Dra & 109387 & B6IIIe & 4.8 & 53.4$\pm$4.7 & 62.9$\pm$6.6 & -\\   
$\chi$ Oph & 148184 & B2Vne & 11.0 & 48.5$\pm$14.9 &  46.7$\pm$5.2 & 23$\pm11$\\   
$\bf{\upsilon}$ Cyg & 202904 & B2Vne & \bf{9.3} &  27.3$\pm$7.8 &  50.0$\pm$13 & 38$\pm9$\\   
o Aqr & 209409 & B7IVe & 4.2 & 75.5$\pm$5.1 &  77.5$\pm$4.6 & 73$\pm18$\\   
$\beta$ Psc & 217891 & B6Ve & 4.7 & 35.9$\pm$7.3 &  32.5$\pm$5.0 & 26$\pm10$\\  \hline
\end{tabular}
\caption{Stellar, interferometric, $\rm H\,\alpha$ profile fitting, and gravity darkening characteristics for the 11 Be stars in the NPOI sample. Note that $\kappa$ Dra lacks an associated value of $i_{\rm GD}$. Additional information on $i_{\rm NPOI}$ and $i_{\rm H\,\alpha}$ can be found in \citet{Sigut2020}; additional information on $i_{\rm GD}$ can be found in \citet{Zorec2016}.}
\label{NPOI_properties}
\end{table}        

The three machine learning algorithms, trained on synthetic \ha\ profiles, were tested on the NPOI sample of observed profiles and the results are compared with the inclination angle determinations made using both interferometry and \ha\ profile fitting. As before, the performance of an algorithm is taken to be the RMSE between its determinations of $i$ and either $i_{\rm NPOI}$ or \iha\ (considered separately). 

\begin{figure*}
\centering
\includegraphics[width=0.97\textwidth ]{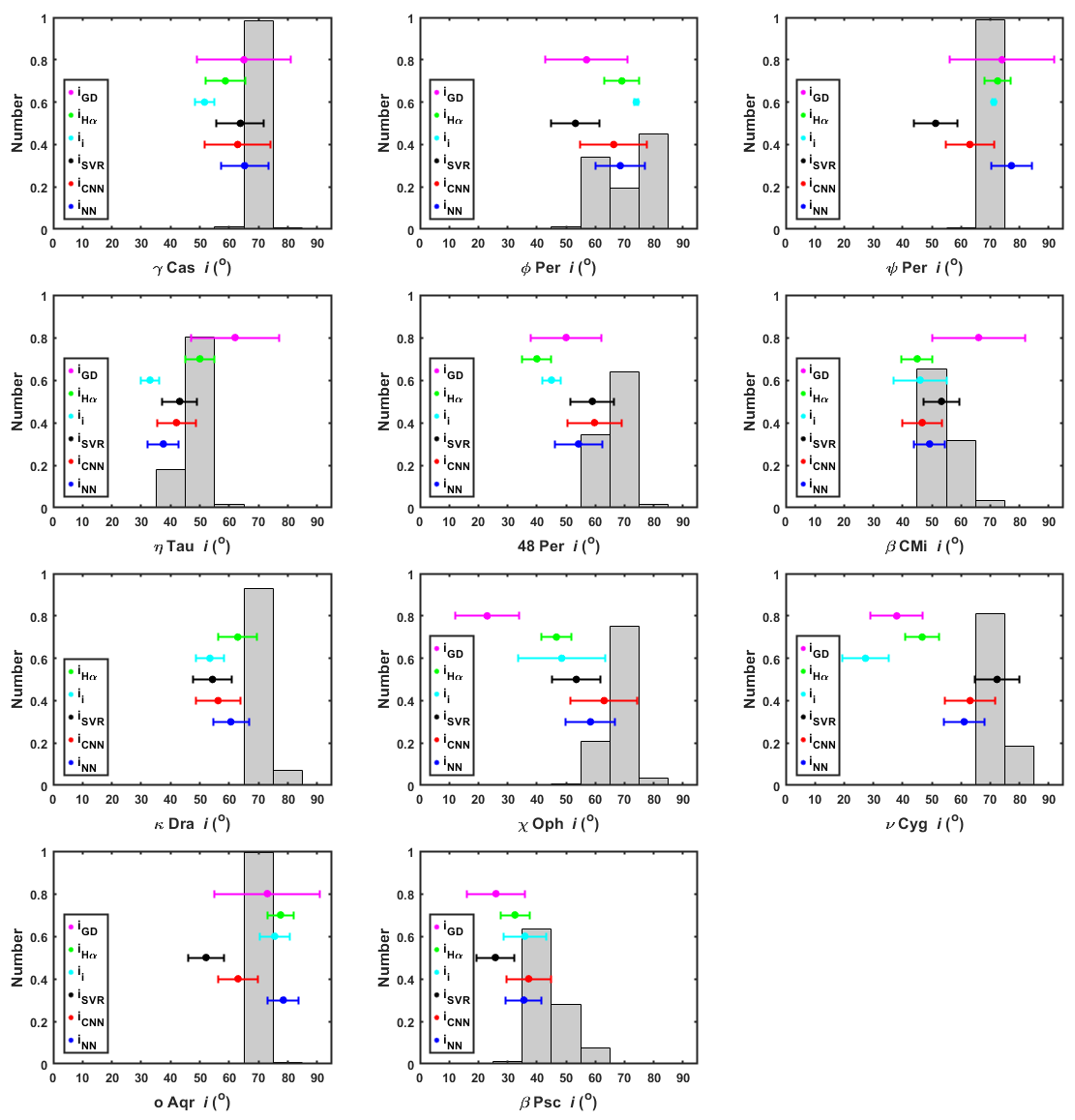}
\caption{Panel plot of the 11 stars of the NPOI sample with inclination determinations from \igd\ (magenta), \iha\ (green), $i_{\rm NPOI}$ (cyan), $i_{\rm SVR}$ (black), $i_{\rm CNN}$ (red), and $i_{\rm NN}$ (blue). Values have been staggered for readability. Also shown in each panel are output histograms of the NNs tasked with classification (shown as grey rectangles whose heights sum to unity) giving the probability assigned to each inclination class. The error bars show $1\sigma$ uncertainties. Note that the values of $i_{\rm CNN}$ are expected to be lower than the centre of their associated histograms because they have been calibrated (see Section~\ref{section:zorec}). Finally, $\kappa\,$Dra does not have a determination of $i_{\rm GD}$.} 
\label{npoi_class_hist}
\end{figure*}

Figure~\ref{npoi_class_hist} shows a comparison of the inclination angle determinations of our three machine learning algorithms  $i_{\rm NN}$, $i_{\rm CNN}$, and $i_{\rm SVR}$ with those of $i_{\rm NPOI}$. The NNs tasked with regression performed the best with an RMSE of $12.3^\circ$. Only two of the 11 determinations of $i$ differed by more than $10^\circ$: $\bf{\upsilon}\,$Cyg ($33.7^\circ$) and $\gamma\,$Cas ($13.6^\circ$). The NNs tasked with classification fared a little worse with a RMSE of $14.2^\circ$. Five of the 11 determinations of $i$ differed by more than $10^\circ$, with the worst cases being those of $\bf{\upsilon}\,$Cyg ($35.8^\circ$) and $\chi\,$Oph ($14.5^\circ$). SVR had the worst performance with a RMSE of $19.0^\circ$. Seven of the 11 determinations of $i$ differed by more than $10^\circ$ with the worst cases being $\bf{\upsilon}\,$Cyg ($43.6^\circ$) and $o\,$Aqr ($24.8^\circ$). Although the inclination determinations of all three algorithms were higher on average than $i_{\rm NPOI}$, the effect was small for SVR ($\mu_{SVR} = +0.4^\circ$) but larger for both types of NNs ($\mu_{NN} = +7.7^\circ, \, \mu_{CNN} = +5.6^\circ$). 

Figure~\ref{npoi_class_hist} also shows a comparison of the inclination angle determinations of $i_{\rm NN}$, $i_{\rm CNN}$, and $i_{\rm SVR}$ with those of \iha. The NNs tasked with regression performed the best with a RMSE of $8.5^\circ$. Four of the 11 determinations of $i$ differed by more than $10^\circ$ with the worst disagreements being $\bf{\upsilon}$ Cyg ($14.3^\circ$) and 48 Per ($14.2^\circ$). The NNs tasked with classification had a RMSE of $11.2^\circ$. Four of the 11 determinations of $i$ differed by more than $10^\circ$ with the worst cases being those of $O$ Aqr ($16.4^\circ$) and $\bf{\upsilon}$ Cyg ($16.3^\circ$). SVR performed the worst with a RMSE of $15.8^\circ$. Five of the 11 determinations of $i$ differed by more than $10^\circ$ with the most discordant cases being $O$ Aqr ($26.8^\circ$) and $\bf{\upsilon}$ Cyg ($24.2^\circ$). 

The relative performance of the three algorithms for the NPOI sample was the same as that of the larger Zorec sample: NNs tasked with regression performed the best, followed by NNs tasked with classification, and then SVR. All three algorithms performed better when compared to \iha\ than they did when compared to \inpoi. This is not surprising because the synthetic profiles used to train the algorithms come from the same libraries as those used for \ha\ profile fitting. 

It is worth highlighting the influence of $\bf{\upsilon}\,$Cyg on the results for the NPOI sample as this star caused all three algorithms significant problems. The three worst discrepancies between an algorithm's determination of $i$ and \inpoi\ all occur for $\bf{\upsilon}\,$Cyg. When comparing an algorithm's determination of $i$ with \iha\, $\bf{\upsilon}\,$Cyg is the largest or second largest discrepancy in all three cases. While $\bf{\upsilon}\,$Cyg does have the smallest value of \inpoi\ in the NPOI sample ($27.3^\circ$), the issue seems to be more complicated than the algorithms struggling with low inclinations because they performed well on both $\eta\,$Tau ($33.0^\circ$) and $\beta\,$Psc ($35.9^\circ$). When comparing with \inpoi, omitting $\bf{\upsilon}\,$Cyg from the sample would cause the following performance changes: NNs tasked with regression would improve by about 40 percent (RMSE falling from $12.3^\circ$ to $7.2^\circ$), NNs tasked with classification to improve by about 30 percent (RMSE falling from $14.2^\circ$ to $9.6^\circ$), and SVR to improve by about 20 percent (RMSE falling from $19.0^\circ$ to $14.5^\circ$). With $\bf{\upsilon}\,$Cyg omitted, these resulting performances are similar to the performances on the full Zorec sample (see Section~\ref{zorec_comparison_iHA}); this may suggest the $\bf{\upsilon}\,$Cyg determinations are anomalous. To resolve whether the inclination angle determinations for $\bf{\upsilon}\,$Cyg really are anomalous, we would ideally like to include more stars in the NPOI sample. Unfortunately, optical interferometry is only possible on the nearest and brightest Be stars, and the question of whether $\bf{\upsilon}\,$Cyg is an anomaly remains unanswered. Finally, when comparing against \iha, the effect of omitting $\bf{\upsilon}\,$Cyg from the NPOI sample results in a smaller performance increase of approximately 10 percent across all three algorithms. 

%
% Aaron -- felt this discussion belaboured things a bit. 
%Conversely, it is also possible that problem stars, such as $\bf{\upsilon}$ Cyg, were underrepresented in the NPOI sample. When comparing with $i_i$, replacing the median performing star with another instance of the performance of $\bf{\upsilon}$ Cyg would cause the performance of: the NNs tasked with regression to decrease by about 30 percent (from $12.3^\circ$ to $15.8^\circ$), the NNs tasked with classification to decrease by about 25 percent (from $14.2^\circ$ to $17.6^\circ$), and SVR to decrease by about 20 percent (from $19.0^\circ$ to $22.8^\circ$). The effects are smaller when comparing against $i_{H\,\alpha}$ and would cause a performance decrease of approximately 10 percent across all three algorithms.   
 
\section{Conclusions}
\label{section:conclusion}

Three supervised machine learning algorithms were trained exclusively on synthetic, Be star \ha\ spectra computed with the \texttt{Bedisk-Beray} code suite to be able to extract an estimate of the central B~star's inclination angle from a single, observed, \ha\ flux profile and the star's spectral type. The algorithms tested were neural networks tasked with regression, neural networks tasked with classification, and support vector regression. When applied to a large ($N\sim 100$) observed sample of Be star spectra \citep{Sigut2022}, neural networks tasked with regression performed best, yielding an inclination accuracy of $\rm RMSE=7.6^\circ$ which is comparable to that obtained by direct model profile fitting of the \ha\ line. Neural networks tasked with classification were an intermediate performer ($\rm RMSE=11^\circ$) and support vector regression performed significantly worse ($\rm RMSE=14^\circ$). 

During the training and hyper-parameter optimizations, it was found that algorithms trained on low $\rm S/N=25$ \ha\ profiles yielded much better results compared to those trained on higher S/N profiles when applied to the real, \ha\ spectra of Be stars. We speculate that the wider variation among the lower S/N synthetic spectra, coupled with the large training samples, allowed the algorithms to better deal with natural variations in observed spectra that are not captured by the models. Training on synthetic data has the advantage that cases rare in the general population (in this case, low inclination systems as $p(i)\,di=\sin i\,di$) can be incorporated into the training, as long as over-specialization of the algorithm to purely synthetic data can be avoided. An interesting avenue for future work is testing how the optimal S/N varies with network depth. Further along these lines, we are testing the viability of training deep, convolutional neural networks on images of \ha\ line profiles (rather than 1D vectors of relative fluxes) to determine the inclination angles of observed Be stars.

Finally, future work will focus on further extending the quantitative analysis of Be star spectra by training neural networks to extract $v\sin i$ estimates from the relevant portions of Be star spectra by focusing on, for example, the observed profiles of He\,{\sc i\;}4471\AA\ and Mg\,{\sc ii\;}4481\AA. We feel that this problem is also very amenable training with synthetic line profiles generated with the \texttt{Bedisk-Beray} code suite. Combined with this future work, the inclination finding neural networks will allow equatorial stellar rotation velocities to be directly measured from moderate-to-high S/N spectra of sufficient resolution. 

\section*{Acknowledgements}

The authors would like to thank the anonymous referee for thoughtful feedback.  B.\ D.\ Lailey acknowledges support from the University of Western Ontario's physics and astronomy department. T.\ A.\ A.\ Sigut acknowledges support, in the form of a Discovery Grant from the Natural Sciences and Engineering Council of Canada.

%%%%%%%%%%%%%%%%%%%%%%%%%%%%%%%%%%%%%%%%%%%%%%%%%%
\section*{Data Availability}

The trained neural networks tasked with regression are available to download at \url{https://github.com/bryanlailey/Be_inclination} and use in the \texttt{MATLAB}~(R2020a or later) programming environment. The observed profiles of the Zorec and NPOI samples and the 4 $\rm M_\odot$ library of synthetic spectra are also available there.  

% The trained neural networks tasked with regression as well as the spectra of the NPOI and Zorec samples are available to download at \url{https://github.com/bryanlailey/Be_inclination}. The libraries of synthetic spectra on which the machine learning algorithms were trained will be made available upon reasonable request.  



\bibliographystyle{mnras}
\bibliography{main} % if your bibtex file is called example.bib

%\begin{thebibliography}{99}
%\end{thebibliography}

%%%%%%%%%%%%%%%%%%%%%%%%%%%%%%%%%%%%%%%%%%%%%%%%%%

%%%%%%%%%%%%%%%%% APPENDICES %%%%%%%%%%%%%%%%%%%%%

%%%%%%%%%%%%%%%%%%%%%%%%%%%%%%%%%%%%%%%%%%%%%%%%%%

% Don't change these lines
\bsp	% typesetting comment
\label{lastpage}
\end{document}